\documentclass[twocolumn,showpacs,preprintnumbers,prb,floatfix,superscriptaddress,amsmath,amssymb]{revtex4}
\usepackage{epsfig}
\usepackage{subfigure}
\usepackage{amsmath}
\usepackage{amssymb}
\usepackage{graphicx}
\usepackage{dcolumn}
\usepackage{bm}
\input epsf

\begin{document}

\title{Synchronization in large directed networks of coupled phase oscillators}

\author{Juan G. Restrepo}
\email{juanga@math.umd.edu} \affiliation{ Institute for Research in
Electronics and Applied Physics, University of Maryland, College
Park, Maryland 20742, USA } \affiliation{ Department of Mathematics,
University of Maryland, College Park, Maryland 20742, USA }

\author{Edward Ott}
\affiliation{ Institute for Research in Electronics and Applied
Physics, University of Maryland, College Park, Maryland 20742, USA }
\affiliation{ Department of Physics and Department of Electrical and
Computer Engineering, University of Maryland, College Park, Maryland
20742, USA }

\author{Brian R. Hunt}
\affiliation{ Department of Mathematics, University of Maryland,
College Park, Maryland 20742, USA }
\affiliation{ Institute for
Physical Science and Technology, University of Maryland, College
Park, Maryland 20742, USA}

\date{\today}

\begin{abstract}
We extend recent theoretical approximations describing the
transition to synchronization in large undirected networks of
coupled phase oscillators to the case of directed networks. We also
consider extensions to networks with mixed positive/negative
coupling strengths. We compare our theory with numerical simulations
and find good agreement.
\end{abstract}


\maketitle

{\bf Synchronization of coupled oscillators is frequently observed in nature and technology
 \cite{pikovsky, mosekilde}.
Recently, the study of synchronization phenomena in complex networks
has received much attention
\cite{pecora2,pecora3,takashi,jadbabaie,bahiana,moreno,ichinomiya,
lee,ichinomiya2,onset}. A classical model for the phase dynamics of
weakly coupled oscillators is that of Kuramoto
\cite{kuramoto,strogatz}, who showed that as the coupling strength
is increased there is a transition from incoherent behavior to
synchronization. The Kuramoto model assumes all-to-all connectivity
and positive coupling (i.e., the coupling of two oscillators tends
to reduce their phase difference). However, it has been recently
noted that the topology of real world networks is often very
complex. In the current paper, generalizing our previous work which
considered the case of undirected coupling networks with positive
coupling\cite{onset}, we discuss the synchronization of phase
oscillators interacting on directed networks with mixed
positive/negative coupling. }

\section{Introduction}

The classical Kuramoto model \cite{kuramoto,strogatz} describes a
collection of globally coupled phase oscillators that exhibits a
transition from incoherence to synchronization as the coupling
strength is increased past a critical value. Since real world
networks typically have a more complex structure than all-to-all
coupling \cite{newman1,barabasi1}, it is natural to ask what effect
interaction structure has on the synchronization transition. In
Ref.~\onlinecite{onset}, we studied the Kuramoto model allowing general
connectivity of the nodes, and found that for a large class of
networks there is still a transition to global synchrony as the
coupling strength exceeds a critical value $k_c$. We found that the
critical coupling strength depends on the largest eigenvalue of the
adjacency matrix $A$ describing the network connectivity. We also
developed several approximations describing the behavior of an order
parameter measuring the coherence past the transition. This past
work was restricted to the case in which $A_{nm}= A_{mn} \geq 0$,
that is, undirected networks in which the coupling tends to reduce
the phase difference of the oscillators.

Most networks considered in applications are directed
\cite{newman1,barabasi1}, which implies an asymmetric adjacency
matrix, $A_{nm} \neq A_{mn}$. Also, in some cases the coupling
between two oscillators might drive them to be out of phase, which
can be represented by allowing the coupling term between these
oscillators to be negative, $A_{nm} < 0$. The effect that the
presence of directed and mixed positive/negative connections can
have on synchronization is, therefore, of interest. Here we show how
our previous theory can be generalized to account for these two
factors. We study examples in which either the asymmetry of the
adjacency matrix or the effect of the negative connections are
particularly severe and compare our theoretical approximations with
numerical solutions.

\section{Background}\label{back}

In Ref.~\onlinecite{onset} we considered the onset of synchronization in networks of heterogeneous
coupled phase oscillators.
This situation can be modeled by the equation,
\begin{equation}\label{eq:coupled}
\dot{\theta}_n = \omega_n + k \sum_{m=1}^{N} A_{nm}\sin(\theta_{m} - \theta_{n}),
\end{equation}
where $\theta_n$, $\omega_n$ are the phase and natural frequency of
oscillator $n$, and $N \gg 1$ is the total number of oscillators.
The frequencies $\omega_n$ are assumed to be independently drawn
from a probability distribution characterized by a density function
$g(\omega)$ that is symmetric about a single local maximum at
$\omega = \overline{\omega}$. The mean frequency $\overline{\omega}$
can be shifted to $\overline{\omega} = 0$ by introduction of the
change of variables $\theta_n \to \theta_n -\overline{\omega} t $.
Thus we henceforth take $\overline{\omega} = 0$. The adjacency
matrix $\{A_{nm}\}$ determines the network connecting the
oscillators. Positive coupling was imposed in Ref.~\onlinecite{onset} by
the condition $A_{nm} \geq 0$. Furthermore, the matrix $A$ was
assumed to be symmetric and thus only undirected networks were
considered. In this Section we will review our results for this
class of networks, following Sec. II of Ref.~\onlinecite{onset}. Thus
throughout this Section $A_{nm}= A_{mn}\geq0$.

In order to quantify the coherence of the inputs to a given node,
a positive real valued local order parameter $r_{n}$ is defined by
\begin{equation}\label{eq:rn}
r_n e^{i \psi_n} \equiv \sum_{m=1}^{N} A_{nm}\langle e^{i \theta_m}\rangle_t,
\end{equation}
where $\langle\dots\rangle_t$ denotes a time average. To characterize the macroscopic
coherence for the whole network,
a global order parameter is defined by
\begin{equation}\label{eq:orderpara}
r = \frac{\sum_{n=1}^{N} r_{n}}{\sum_{n=1}^{N} d_{n}},
\end{equation}
where $d_n$ is the degree of node $n$ defined by
\begin{equation}
d_n = \sum_{m=1}^N A_{nm}.
\end{equation}

In terms of $r_n$, Eq.~(\ref{eq:coupled}) can be rewritten as
\begin{equation}\label{eq:coupled0}
\dot{\theta}_n = \omega_n - k r_{n} \sin(\theta_n - \psi_n) -k h_n(t),
\end{equation}
where the term $h_n(t)$ takes into account time fluctuations and is
given by $h_n = Im \{ e^{-i\theta_n}\sum_m A_{nm}\left( \langle
e^{i\theta_m}\rangle_t - e^{i\theta_m}\right)\} $, where $Im$ stands
for the imaginary part. As argued in Ref.~\onlinecite{onset}, when the number
of connections into each node is large, the term $h_n$ is small
compared to $r_n$ and we obtain approximately
\begin{equation}\label{eq:coupled2}
\dot{\theta}_n = \omega_n - k r_{n} \sin(\theta_n - \psi_n).
\end{equation}
Henceforth, we will assume that the number of connections into each
node is large enough that we can neglect the time fluctuations
represented by the term $h_n$. For a discussion of the effect of
nodes with few connections, see Sec. VI of Ref.~\onlinecite{onset}.

From Eq.~(\ref{eq:coupled2}), we conclude that oscillators with $\left|\omega_n\right| \leq k r_n$
become locked,
i.e., for these oscillators $\theta_n$ settles at a value for which
\begin{equation}\label{eq:locked}
\sin(\theta_{n}-\psi_n) = \omega_n/(k r_n).
\end{equation}
Then
\begin{eqnarray}\label{splitup}
r_n =  \sum_{\left|\omega_{m}\right| \leq k r_{m}} A_{nm}  e^{i(\theta_m - \psi_n)}\\
+   \sum_{\left|\omega_{m}\right| > k r_{m}} A_{nm} \langle e^{i(\theta_m - \psi_n)}\rangle_t.\nonumber
\end{eqnarray}
The sum over the non-locked oscillators can be shown to vanish in the large number of connections per node
 limit (see Appendix A of Ref.~\onlinecite{onset}),
and we obtain from the real and imaginary parts of
Eq.~(\ref{splitup})
\begin{eqnarray}\label{eq:betaprime}
r_n = \sum_{\left|\omega_{m}\right| \leq k r_{m} }
 A_{nm} \cos(\psi_m - \psi_n)\sqrt{1 - \left(\frac{\omega_{m}}{k r_{m}}\right)^2}\\
- \sum_{\left|\omega_{m}\right| \leq k r_{m} }
 A_{nm} \sin(\psi_m - \psi_n)\left(\frac{\omega_{m}}{k r_{m}}\right),\nonumber
\end{eqnarray}
and
\begin{eqnarray}\label{eq:imaginary}
0 = \sum_{\left|\omega_{m}\right| \leq k r_{m}  }
 A_{nm} \cos(\psi_m - \psi_n) \left(\frac{\omega_{m}}{k r_{m}}\right)\\
+ \sum_{\left|\omega_{m}\right| \leq k r_{m}  }
 A_{nm} \sin(\psi_m - \psi_n)\sqrt{1-\left(\frac{\omega_{m}}{k r_{m}}\right)^2}.\nonumber
\end{eqnarray}
Introducing the assumption that the solutions $\psi_n$, $r_n$ are statistically
independent of $\omega_n$ (see Ref.~\onlinecite{onset}) and using the assumed symmetry of the frequency
distribution $g(\omega)$
we obtain from Eq.~(\ref{eq:betaprime}) the approximation,
\begin{equation}\label{eq:cosfi}
r_n = \sum_{\left|\omega_{m}\right| \leq k r_{m} }
 A_{nm} \cos(\psi_m - \psi_n)\sqrt{1 - \left(\frac{\omega_{m}}{k r_{m}}\right)^2},
\end{equation}
and the right side of Eq.~(\ref{eq:imaginary}) is approximately zero
for large number of connections per node. The solution of
Eq.~(\ref{eq:cosfi}) with $\psi_n = \psi_m$ for all $n$ is the one
corresponding to the smallest value of $k$, and thus corresponds to
the smallest critical coupling $k_{c}$ leading to a transition to a
macroscopic value of $r_n$. Therefore we consider the equation
\begin{equation}\label{eq:betass}
r_n = \sum_{\left|\omega_{m}\right| \leq k r_{m} }
 A_{nm} \sqrt{1 - \left(\frac{\omega_{m}}{k r_{m}}\right)^2}.
\end{equation}
We refer to this approximation [Eq.~(\ref{eq:betass})], based on neglecting the time fluctuations,
as the {\it time averaged theory} (TAT). In Ref.~\onlinecite{onset} we showed numerically that
this approximation consistently describes the large time behavior of the order parameter $r$ past
the transition for
various undirected networks with positive coupling strengths (i.e., $A_{nm} = A_{mn} \geq 0$).

Averaging over the frequencies, one obtains the {\it frequency distribution approximation} (FDA):
\begin{equation}\label{eq:betaint}
r_{n} = k {\sum_{m}} A_{nm} r_{m} \int_{-1}^{1} g(z k r_{m}) \sqrt{1 - z^2 } dz.
\end{equation}

The value of the critical coupling strength can be obtained from the frequency distribution
 approximation by letting
$r_n \to 0^+$, producing
\begin{equation}\label{eq:firstor}
r_{n}^{(0)} = \frac{k}{k_0} {\sum_{m}} A_{nm} r_{m}^{(0)},
\end{equation}
where $k_{0} \equiv 2/[\pi g(0)]$. The critical coupling strength thus corresponds to
\begin{equation}\label{eq:kc}
k_{c} = \frac{k_0}{\lambda},
\end{equation}
where $\lambda$ is the largest eigenvalue of the adjacency matrix
$A$ and $r^{(0)}$ is proportional to the corresponding eigenvector
of $A$. By considering perturbations from the critical values as
$r_n = r_n^{(0)} + \delta r_n$, expanding $g(z k r_m)$ in
Eq.~(\ref{eq:betaint}) to second order  for small argument,
multiplying Eq.~(\ref{eq:betaint}) by $r_n^{(0)}$ and summing over
$n$, we obtained an expression for the order parameter past the
transition valid for networks with relatively homogeneous degree
distributions \cite{footnote}:
\begin{equation}\label{perturba}
r^2 = \left(\frac{\eta_{1}}{\alpha k_{0}^2}\right)
\left(\frac{k}{k_{c}} - 1\right)
\left(\frac{k}{k_{c}}\right)^{-3}
\end{equation}
for $0< (k/k_c) -1\ll 1$, where
\begin{equation}\label{eq:eta1}
\eta_1 \equiv \frac{\langle u\rangle^2 \lambda^2}{ N \langle d\rangle^2 \langle u^4\rangle},
\end{equation}
$\alpha = -\pi g''(0)k_0/16$, $u$ is the normalized eigenvector of $A$ corresponding to $\lambda$,
and $\langle\dots\rangle$ is defined by $\langle x^q\rangle = \sum_{n = 1}^N x^q_n / N$.

The {\it mean field theory} (MFT) \cite{ichinomiya,lee} was obtained from the frequency
distribution equation by introducing the extra
assumption that the local mean field is approximately proportional to the degree, $r_n = r d_n$.
Substituting this into Eq.~(\ref{eq:betaint}) and summing over $n$ we obtained
\begin{equation}\label{eq:betasumed}
\sum_{m = 1}^N d_{m} = k \sum_{m = 1}^N d_{m}^2 \int_{-1}^{1} g(z k r d_{m}) \sqrt{1 - z^2 } dz.
\end{equation}
Letting $r\to 0^+$, the critical coupling strength is given by
\begin{equation}\label{eq:firstorder}
k \equiv k_{mf} = k_{0} \frac{\langle d \rangle}{\langle d^2 \rangle}.
\end{equation}
An expansion to second order yields
\begin{equation}\label{eq:secondmf}
r^2 =  \left(\frac{\eta_{2}}{\alpha k_{0}^2}\right)
\left(\frac{k}{k_{mf}} - 1\right)\left( \frac{k}{k_{mf}}\right)^{-3}
\end{equation}
for $0< (k/k_{mf}) -1\ll 1$, where
\begin{equation}\label{eq:eta2}
\eta_{2} \equiv \frac{\langle d^2\rangle^3}{\langle d^4\rangle \langle d \rangle^2}.
\end{equation}

Comparing the above three approximations, we note the following points:
\begin{enumerate}
\item The TAT requires knowledge of the adjacency matrix and the particular realization of the oscillator
frequencies $\omega_n$ at each node.
\item The FDA requires knowledge of the adjacency matrix and the frequency distribution,
but averages over realizations
of the node frequencies.
\item The MFT (like the FDA) averages over realizations of the node frequencies, but
only requires knowledge of the degree distribution $d_m$ (knowledge
of the  adjacency matrix is not required).
\item Computationally, the TAT and
the FDA are more demanding than
the MFT; all three, however, are much less costly than direct
integration of Eq.~(\ref{eq:coupled}) to find the time asymptotic
result.
\item Finally, one might suspect that the TAT is more accurate for describing a specific system realization, given that
one has knowledge of the network and the realization of the oscillator frequencies $\omega_n$ on each node, while the FDA
might be more appropriate for investigating the mean behavior averaged over an ensemble of realizations of the
oscillator frequencies.
\end{enumerate}

\section{Directed networks}\label{dire}

In this Section we will extend our previous results to include
directed networks, $A_{nm} \neq A_{mn}$. As in the previous Section,
we will assume that the number of connections per node (both
incoming and outgoing) is large, that the frequencies are drawn
randomly from a distribution symmetric around its unique local
maximum at $\omega = 0$, and that the coupling is positive, $A_{nm}
\geq 0$. We define the {\it in-degree} $d_n^{in}$ and {\it
out-degree} $d_n^{out}$ of node $n$ as
\begin{equation}
d_n^{in}\equiv \sum_{m=1}^N A_{nm}
\end{equation}
and
\begin{equation}
d_n^{out}\equiv \sum_{m=1}^N A_{mn}.
\end{equation}
For directed networks, the degrees $d_n^{in}$ and $d_n^{out}$ may be
unequal, and  it is therefore necessary to take this difference into
account when developing approximations for the synchronization
transition based on the degree of the nodes [e.g., the mean field
theory, Eq.~(\ref{eq:betasumed})].

The approximations to $r$ given by the time averaged theory
[Eq.~(\ref{eq:betass})], the frequency distribution approximation
[Eq.~(\ref{eq:betaint})], and the estimate for the critical coupling
constant given by Eq.~(\ref{eq:kc}) are still valid in this more
general case. The existence of a nonnegative real eigenvalue
$\lambda$ larger than the magnitude of any other eigenvalue is
guaranteed for matrices with nonnegative entries by the Frobenius
theorem \cite{bapat}, and we use this eigenvalue in
Eq.~(\ref{eq:kc}).

We now consider the perturbation solution to the FDA
[Eq.~(\ref{eq:betaint})] for $(k-k_c)$ small taking into account
asymmetry of $A$. Expanding Eq.~(\ref{eq:betaint}) to second order
in $k r_n$, inserting $r_n = r_n^{(0)} + \delta r_n$, and canceling
terms of order $r_n^{(0)}$, the leading order terms remaining are
\begin{eqnarray}\label{eq:fea}
\delta r_n = \frac{k}{k_{c}\lambda} \sum_{m} A_{nm}\delta r_m -
\frac{\alpha k^3}{k_{c}\lambda}\sum_{m} A_{nm} (r_m^{(0)})^3 \\
+ \frac{k - k_{c}}{k_c \lambda} \sum_{m} A_{nm} r_m^{(0)}\nonumber.
\end{eqnarray}
In order for Eq.~(\ref{eq:fea}) to have a solution for $\delta r_n$,
it must satisfy a solubility condition. This condition can be
obtained as follows. Let $\overline{u}_n$ be an eigenvector of the
transpose of $A$, $A^{T}$, with eigenvalue $\lambda$. Multiplying
Eq.~(\ref{eq:fea}) by $\overline{u}_n$, summing over $n$ and using
Eq.~(\ref{eq:firstor}), we obtain
\begin{equation}
\frac{\sum_{m} (r_m^{(0)})^3 \overline{u}_m}{\sum_{m} r_m^{(0)}\overline{u}_m} = \frac{k - k_c}{\alpha k^3}.
\end{equation}
In terms of $u$ and $\overline{u}$, eigenvectors of $A$ and $A^T$
associated with the eigenvalue $\lambda$, the square of the order
parameter $r$ can be expressed as [cf. Eqs.~(\ref{perturba}) and
(\ref{eq:eta1})]
\begin{equation}\label{eq:secondpt}
r^2 = \left(\frac{\overline{\eta}_{1}}{\alpha k_{0}^2}\right)
\left(\frac{k}{k_{c}} - 1\right)
\left(\frac{k}{k_{c}}\right)^{-3}
\end{equation}
for $0< (k/k_c) -1\ll 1$, where
\begin{equation}\label{eq:eta1dire}
\overline{\eta}_1 \equiv \frac{\langle u\rangle^2 \langle u \overline{u}\rangle  \lambda^2 }{ N \langle d\rangle^2
\langle u^3 \overline{u}\rangle},
\end{equation}
and $\langle x^p y^q \rangle$ is defined by $\langle x^p y^q \rangle
= \sum_{n=1}^N x_n^p y_n^q/N$. We will refer to this generalization
of the perturbation theory as the {\it directed perturbation theory}
(DPT).

The mean field theory can also be generalized for directed networks
by introducing the assumption $r_n = r d_n^{in}$. We obtain as a
generalization of Eq.~(\ref{eq:betasumed}) the {\it directed mean
field theory} (DMFT)

\begin{equation}\label{mftdire}
\sum_{m = 1}^N d_{m}^{in} = k \sum_{m = 1}^N d_{m}^{in}d_m^{out} \int_{-1}^{1} g(z k r d_{m}^{in}) \sqrt{1 - z^2 } dz.
\end{equation}
Letting $r\to 0^+$, the critical coupling strength is given by
\begin{equation}\label{eq:firstorder2}
k \equiv k_{mf} = k_{0} \frac{\langle d^{in} \rangle}{\langle d^{in}d^{out} \rangle}.
\end{equation}
An expansion to second order yields [cf. Equations~(\ref{eq:secondmf}) and (\ref{eq:eta2})]
\begin{equation}\label{eq:secondmfdire}
r^2 =  \left(\frac{\overline{\eta}_{2}}{\alpha k_{0}^2}\right)
\left(\frac{k}{k_{mf}} - 1\right)\left( \frac{k}{k_{mf}}\right)^{-3}
\end{equation}
for $0< (k/k_{mf}) -1\ll 1$, where
\begin{equation}\label{eq:eta2dire}
\overline{\eta}_{2} \equiv \frac{\langle d^{in} d^{out}\rangle^3}{\langle (d^{in})^3 d^{out}\rangle \langle d^{in} \rangle^2}.
\end{equation}

\section{Networks with negative coupling}\label{inhibitory}

Here we extend our previous results to the case in which the matrix
elements $A_{nm}$ are allowed to be negative. In this case, a
solution to Eqs.~(\ref{eq:betaprime}) and (\ref{eq:imaginary}) in
which all the phases are equal, ($\psi_n = \psi_m$ for all $n$,$m$),
does not necessarily exist. [In fact, if one were to set $\psi_n =
\psi_m$ in Eq.~(\ref{eq:cosfi}) the right hand side of
Eq.~(\ref{eq:betass}) could be negative, while by definition $r_n$
is nonnegative.]

Although in this section we will assume $k \geq 0$, the case $k < 0$
can be treated by redefining $k \to -k$ and $A_{nm} \to -A_{nm}$. By
neglecting the contribution of the drifting oscillators, using the
symmetry of $g(\omega)$ and the assumed independence of $\psi_n$ and
$r_n$ from $\omega_n$, we obtain from Eqs.~(\ref{eq:rn}),
(\ref{eq:locked}) and (\ref{splitup}) the equation
\begin{eqnarray}\label{eq:betaprime2}
r_n e^{i\psi_n} = \sum_{\left|\omega_{m}\right| \leq k r_m}A_{nm} e^{i\psi_m} \sqrt{ 1 - \left(\frac{\omega_{m}}{k r_m}\right)^2  }.
\end{eqnarray}
Our approach will now be to solve Eq.~(\ref{eq:betaprime2}) numerically for $\psi_n$ and $r_n$.
We note that such numerical solution will still be orders of magnitude faster than finding the
exact temporal evolution of the network by numerically integrating Eqs.~(\ref{eq:coupled}).
In order to numerically solve Eq.~(\ref{eq:betaprime2})
for the variables $\psi_n$, $r_n$, we look for fixed points
of the following mapping, $(r_n^{j},\psi_n^{j})\to (r_n^{j+1},\psi_n^{j+1})$, defined by
\begin{eqnarray}\label{eq:betaprime3}
r_n^{j+1} e^{i\psi_n^{j+1}} = \sum_{\left|\omega_{m}\right| \leq k r_m^{j}}
A_{nm} e^{i\psi_m^{j}} \sqrt{ 1 - \left(\frac{\omega_{m}}{k r_m^{j}}\right)^2  }.
\end{eqnarray}
Repeatedly iterating the above map starting from random initial
conditions, the desired solution will be produced if the orbit
converges to a fixed point. We will discuss the convergence of this
procedure when considering particular examples.

We now comment on some aspects introduced by connections with
negative coupling. First, we note that when the coupling between the
oscillators is positive, the effect of the coupling between them is
a tendency to reduce their phase difference. In this case, as $k\to
\infty$, the phases synchronize, $\theta_n \to 0$. There is in this
case frequency and phase synchronization [i.e.,
$\frac{d}{dt}(\theta_n - \theta_m)\to 0$ and $(\theta_n -
\theta_m)\to 0$]. On the other hand, two oscillators coupled with a
negative connection $A_{nm} < 0$ tend to oscillate out of phase.
However, in a network with many nodes and mixed positive/negative
connections, the relative phases of two oscillators can not in
general be determined only from the sign of their coupling. When the
oscillators lock, their relative phase is determined by $\psi_n$
[let $k \to \infty$ in Eq.~(\ref{eq:locked})], and in general the
phases $\psi_n$ can be broadly distributed in $[0,2\pi)$. Therefore
in this case we expect frequency synchronization, but not phase
synchronization [i.e., $\frac{d}{dt}(\theta_n - \theta_m)\to 0$ but
$(\theta_n - \theta_m)\nrightarrow 0$]. We also note that in this
case the order parameter $r$, as we have defined it in
Eq.~(\ref{eq:orderpara}), may attain values higher than $1$ for
$k\to \infty$. We therefore replace the definition
(\ref{eq:orderpara}) by
\begin{equation}\label{eq:orderpara2}
r = \frac{\sum_{n=1}^{N} r_{n}}{\sum_{m=1}^{N}\sum_{n=1}^{N} |A_{nm}|}.
\end{equation}
Note that if $A_{nm} \geq 0$ this definition reduces to the previous one.

From Eq.~(\ref{eq:cosfi}) we have for $k\to \infty$
\begin{equation}\label{eq:cosfi2}
r\to \frac{\sum_{m,n}A_{nm}\cos(\psi_m-\psi_n)}{\sum_{m,n}|A_{nm}|}.
\end{equation}
The order parameter achieves its maximum value, $r = 1$, when the
phase difference $\psi_m - \psi_n$ between two oscillators is $0$
for positive coupling ($A_{nm} > 0$) and $\pi$ for negative coupling
($A_{nm} < 0$). An order parameter smaller than $1$ as $k\to \infty$
indicates frustration in the collection of coupled oscillators,
i.e., the phase difference favored by the coupling between each pair
of oscillators cannot be satisfied simultaneously by all pairs
\cite{daido2}. The order parameter is similar to the overlap
function used in neural networks for measuring the closeness of the
state of the network to a memorized pattern \cite{takashi2}.

Using the assumption that the number of connections per node is
large, we average Eq.~(\ref{eq:betaprime2}) over the frequencies to
obtain the approximation
\begin{eqnarray}\label{eq:fdainhi}
r_n e^{i \psi_n} = k \sum_{m=1}^N
A_{nm} e^{i\psi_m} r_m \int_{-1}^{1}\sqrt{1 - z^2}g(z k r_m)dz.
\end{eqnarray}
The critical coupling strength $k_c$ can be estimated by letting
$r_n \to 0^+$ to be as in Sec.~\ref{back}
\begin{equation}\label{kcinhi}
k_c = \frac{k_0}{\lambda},
\end{equation}
where $k_0=2/[\pi g(0)]$ and we have assumed the existence of a
positive real eigenvalue $\lambda$ which is larger than the real
part of all other (possibly complex) eigenvalues of $A$. We now
discuss the validity of this assumption.

If the adjacency matrix $A$ is asymmetric and there are mixed
positive/negative connections (both $A_{nm} > 0$ and $A_{n'm'} < 0$
for some $n$,$m$,$n'$,$m'$), it might occur that the matrix $A$ has
no real eigenvalues, or it has complex eigenvalues with real part
larger than the largest real eigenvalue. In our examples we find,
however, that when there is a bias towards positive coupling
strengths, there is a real eigenvalue $\lambda$ with real part
larger than that of the other eigenvalues. Furthermore, the largest
real part of the remaining eigenvalues is typically well separated
from $\lambda$. This issue is discussed further and illustrated with
the spectrum of a particular matrix in Appendix \ref{appendixa}.

So far, we have considered situations in which coupling from
oscillator $m$ to oscillator $n$ favors a phase difference $\theta_n
- \theta_m = 0$ (positive coupling, $A_{nm} >0$), or situations in
which a phase difference $\theta_n - \theta_m = \pi$ is favored
(negative coupling, $A_{nm} < 0$). A more general case is that in
which coupling from oscillator $m$ to oscillator $n$ favors a phase
difference $\theta_n - \theta_m = \alpha_{nm}$, with $0\leq
\alpha_{nm} < 2\pi$. (Such nontrivial phase differences could be
favored, for example, by a time delay in the interaction of the
oscillators in conditions in which, in the absence of a delay, their
interaction would reduce their phase difference to zero.) This more
general case can be described by the following generalization of
Eq.~(\ref{eq:coupled}):
\begin{equation}\label{eq:alphas}
\dot{\theta}_n = \omega_n + k \sum_{m=1}^{N} |A_{nm}|\sin(\theta_{m}
- \theta_{n} + \alpha_{nm}).
\end{equation}
In this scenario, positive coupling corresponds to $\alpha_{nm} = 0$
and negative coupling to $\alpha_{nm} = \pi$.
By considering complex values of the coupling constants,
\begin{equation}
A_{nm} = |A_{nm}|e^{i\alpha_{nm}},
\end{equation}
the same process described at the beginning of this Section can be
used to show that Eq.~(\ref{eq:betaprime2}) is still valid in this
more general case. For simplicity, in our examples we will consider
cases in which $\alpha_{nm}$ is either $0$ or $\pi$.

\section{Examples}\label{examples}

In this section we will numerically test our approximations (Secs. \ref{dire} and \ref{inhibitory})
with examples.

In Ref.~\onlinecite{onset} we showed how our theory described the behavior
of the order parameter $r$ for a particular realization of the
network and the frequencies. Although the agreement was very good,
there was a small but noticeable difference between the time
averaged theory and the frequency distribution approximation. Here,
besides the asymmetry of the adjacency matrix, we will investigate
the variations that occur when different realizations of the network
and the frequencies of the individual oscillators are considered. We
will show that the small discrepancies mentioned above can be
accounted for by averaging over many realizations of the
frequencies.

We will compare the approximations described in this section with
the numerical solution of Eq.~(\ref{eq:coupled}) for different types
of networks. When numerically solving Eq.~(\ref{eq:coupled}), the
initial conditions for $\theta_n$ are chosen randomly in the
interval $[0,2\pi)$ and Eq.~(\ref{eq:coupled}) is integrated forward
in time until a stationary state is reached (stationary state here
means stationary in a statistical sense; i.e., although the solution
might be time dependent, its statistical properties remain constant
in time). From the values of $\theta_n(t)$ obtained for a given $k$,
the order parameter $r$ is estimated using Eqs.~(\ref{eq:rn}) and
(\ref{eq:orderpara}), where the time average is taken after the
system reaches the stationary state. (Close to the transition, the
time needed to reach the stationary state is very long, so that it
is difficult to estimate the real value of $r$. This problem also
exists in the classical Kuramoto all-to-all model.) The value of $k$
is then increased and the system is allowed to relax to a stationary
state, and the process is repeated for increasing values of $k$.
Throughout this section, the frequency distribution is taken to be
$g(\omega) = \frac{3}{4}(1 - \omega^2)$ for $\left|\omega\right|
\leq 1$ and $0$ otherwise.

\subsection{Example (i), A Randomly Asymmetric Network with $A_{nm} > 0$}

As our first example [example (i)] we consider a directed random
network generated as follows. Starting with $N \gg 1$ nodes, we
consider all possible ordered pairs of nodes $(n,m)$ with $n \neq m$
and add a directed link from node $n$ to node $m$ with probability
$s$. (Equivalently, each nondiagonal entry of the adjacency matrix
is independently chosen to be $1$ with probability $s$ and $0$ with
probability $1-s$, and the diagonal elements are set to zero.) Even
though the network constructed in this way is directed, for most
nodes $d_n^{in} \approx d_n^{out}$.
\begin{figure}[h]
\begin{center}
\epsfig{file = 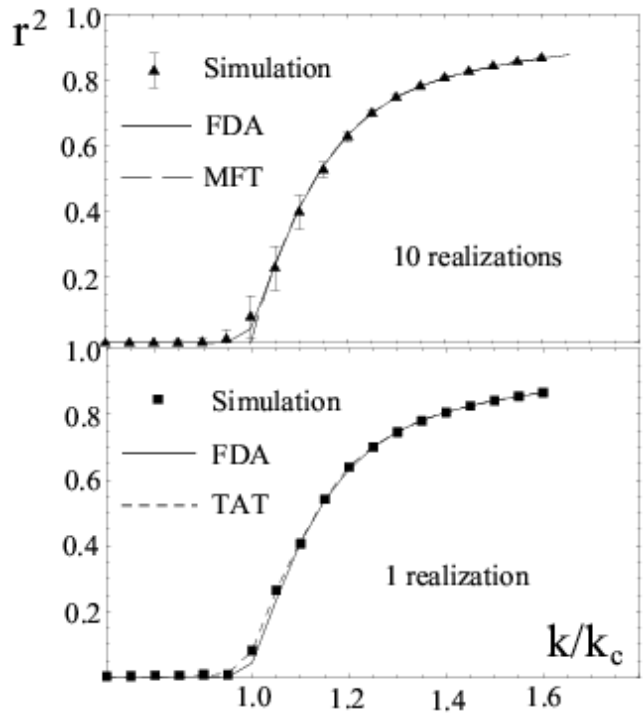, clip =  ,width=1.0\linewidth}
\caption{
(a) Average of the order parameter $r^2$ obtained from numerical solution of Eq.~(\ref{eq:coupled})
 over $10$ realizations of the
network and frequencies (triangles), from the frequency distribution
approximation (solid line) and from the directed mean field theory
(long dashed line) as a function of $k/k_c$.
(b) Order parameter
$r^2$ obtained from numerical solution of Eq.~(\ref{eq:coupled}) for
a particular realization of the network and frequencies (boxes),
from the time averaged theory (short dashed line) and from the
frequency distribution approximation (solid line) as a function of
$k/k_c$. } \label{fig:figpeasy}
\end{center}
\end{figure}
For $N = 1500$ and $s = 2/15$, Fig.~\ref{fig:figpeasy}(a) shows the
average of the order parameter  $r^2$ obtained from numerical
solution of Eq.~(\ref{eq:coupled}) averaged over $10$ realizations
of the network and frequencies (triangles), the  frequency
distribution approximation (FDA, solid line), and the mean field
theory (MFT, long dashed line) as a function of $k/k_c$, where the
results for the FDA and the MFT are averaged over the $10$ network
realizations (note, however, that the FDA and the MFT do not depend
on the frequency realizations). (The perturbation theory
Eq.~(\ref{perturba}) agreed with the frequency distribution
approximation and was left out for clarity.) The error bars
correspond to one standard deviation of the sample of $10$
realizations. We note that the larger error bars occur after the
transition. When the values of the order parameter are averaged over
$10$ realizations of the network and the frequencies, the results
show very good agreement with the frequency distribution
approximation and the directed mean field theory.

In order to study how well our theory describes single realizations,
we show in Fig.~\ref{fig:figpeasy}(b) the order parameter $r^2$
obtained from numerical solution of Eq.~(\ref{eq:coupled}) for a
particular realization of the network and frequencies (boxes), the
time averaged theory (short dashed line), and the frequency
distribution approximation (solid line) as a function of $k/k_c$. As
can be observed from the figure, in contrast with the time averaged
theory, the frequency distribution approximation deviates from the
numerical solution (boxes) by a small but noticeable amount. This
behavior is observed for the other realizations as well. We note
that the FDA and MFT results are virtually identical for all $10$
realizations. On the other hand, the TAT and the results from direct
numerical solution of Eq.~(\ref{eq:coupled}) show dependence on the
realization. Since the FDA and MFT incorporate the realizations of
the connections $A_{nm}$, but not the frequencies, we interpret the
observed realization dependence of the TAT and the direct solutions
of Eq.~(\ref{eq:coupled}) as indicating that the latter dependence
is due primarily to fluctuations in the realizations of the
frequencies rather than to fluctuations in the realizations of
$A_{nm}$.

Note that for our example $N = 1500$ and $s = 2/15$ implies that on
average we have $d^{in} \approx d^{out} \approx 200$. Thus for
comparison purposes, we generated an undirected network as follows:
starting with $N = 1500$ nodes, we join pairs of nodes with
undirected links in such a way that all nodes have $d_n^{in}=
d_n^{out} = 200$. This is accomplished by using the configuration
model described in Sec. IV of Ref.~\onlinecite{newman1}. The resulting
network is described by a symmetric adjacency matrix $A$. The
results for this network are similar to those shown in the previous
example. This suggests that the asymmetric network in the previous
example can be considered (in a statistical sense) as symmetric.

In summary, for the  random asymmetric network in example (i) and
for the symmetric network described in the previous paragraph (not
shown), all the approximations work satisfactorily: single
realizations are described by the time averaged theory, and the
average over many realizations is described by the frequency
distribution approximation or the directed mean field theory.

\subsection{Example (ii), A Strongly Asymmetric Network with $A_{nm} > 0$}

Now we consider a network in which the asymmetry has a more
pronounced effect [example (ii)]. We consider directed networks
defined in the following way. Using the configuration model as
above, we first randomly generate an undirected network with $N =
1500$ nodes and $400$ connections to each node, obtaining a
symmetric adjacency matrix $A'$ with entries $0$ or $1$. We
construct directed networks from this undirected network as follows.
From the symmetric matrix $A'$, $1$'s above the diagonal are
independently converted into $0$'s with probability $1-p$,
generating by this process an asymmetric adjacency matrix $A$.
(Imagining that the nodes are arranged in order of ascending $n$
along a line, connections pointing in the direction of increasing
$n$ are randomly removed. This could model, for example, oscillators
which are coupled chemically along the flow of some medium, or
flashing fireflies that are looking mostly in one direction.) We
will consider a rather low value of $p$, $p = 0.1$, in order to
obtain a network with a strong asymmetry.

\begin{figure}[h]
\begin{center}
\epsfig{file = 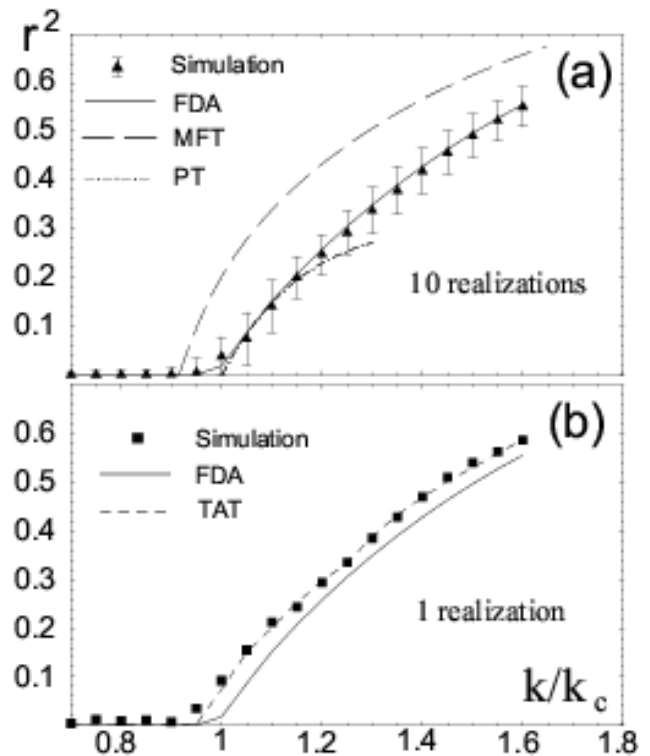, clip =  ,width=1.0\linewidth}
\caption{
(a) Average of the order parameter $r^2$ obtained from numerical solution of
 Eq.~(\ref{eq:coupled}) over $10$ realizations of the
network and frequencies with $p = 0.1$ (triangles), from the frequency
distribution approximation (solid line),
from the directed mean field theory (long dashed line), and from the directed perturbation
theory (dotted-dashed line) as a function of $k/k_c$.
(b)
Order parameter $r^2$ obtained from numerical solution of Eq.~(\ref{eq:coupled})
for a particular realization of the
network and frequencies (boxes), from the time averaged theory (short dashed line) and
from the frequency distribution approximation (solid line) as a function of $k/k_c$.}
\label{fig:figpe110}
\end{center}
\end{figure}

In Fig~\ref{fig:figpe110} we compare our approximations against
the values of the order parameter obtained from
numerical solution of Eq.~(\ref{eq:coupled}) as a function of $k/k_c$
for a network constructed as described above where $k_c$ is given by Eq.~(\ref{eq:kc}).
In Fig.~\ref{fig:figpe110}(a) we show the average of the order
parameter $r^2$ [defined by Eq.~(\ref{eq:orderpara})] versus $k/k_c$
obtained from numerical solution of Eq.~(\ref{eq:coupled}) over $10$
realizations of the network and frequencies (triangles), the
frequency distribution approximation (solid line), the directed mean
field theory Eq.~(\ref{mftdire}) (long dashed line) and the directed
perturbation theory Eq.~(\ref{eq:secondpt}) (dotted-dashed line).
The frequency distribution approximation captures, as in the
undirected case, the values of the average of the order parameter
obtained from numerical solution of Eq.~(\ref{eq:coupled}). The
directed perturbation theory gives a good approximation for small
values of $k$ close to $k_c$, as expected. On the other hand, the
directed mean field theory predicts a transition point which is
smaller than the one actually observed. We note that for this
network solutions of Eq.~(\ref{eq:coupled}) yield substantial rms
deviation of individual realizations [the error bars in
Fig.~\ref{fig:figpe110}(a)] for all $k > k_c$.

Now we consider a single realization. In Fig.~\ref{fig:figpe110}(b) we show the order
 parameter $r^2$ obtained from numerical
solution of Eq.~(\ref{eq:coupled}) for a particular realization of
the network and frequencies (boxes), the time averaged theory (short
dashed line) and the frequency distribution approximation (solid
line) as a function of $k/k_c$. The time averaged theory tracks the
value of the order parameter for this particular realization. This
is also observed for the other realizations.

As an indication of why the directed mean field theory gives
a smaller transition point than that given by $k_c$ in Eq.~(\ref{eq:kc}),
we note that in the limiting case,
$p \to 0$, all the elements above and in the diagonal of $A$ are $0$, so that $\lambda = 0$
and $k_c \to \infty$.
However, the directed mean field theory predicts a transition at the finite value
$k_{mf} = k_{0} \langle d^{in}\rangle/(\langle d^{in}d^{out}\rangle)$.

\begin{figure}[h]
\begin{center}
\epsfig{file = 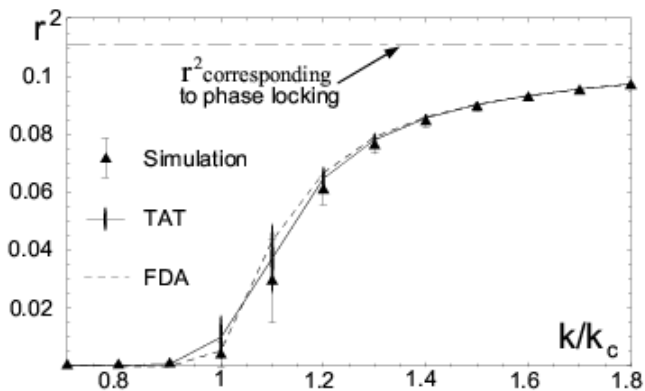, clip =  ,width=1.0\linewidth}
\caption{
Average of the order parameter $r^2$ obtained from numerical solution of
Eq.~(\ref{eq:coupled})
over $10$ realizations of the network with $q = 2/3$ and frequencies
(triangles with thin error bars),
average of the time averaged theory (solid line with oval error bars), and
frequency distribution approximation (dashed line) as a function of $k/k_c$.
The horizontal line represents the value of the order parameter if the
oscillators were phase locked ($\theta_n = \theta_m$ for all $m$ and $n$).}
\label{inhi}
\end{center}
\end{figure}

\subsection{Examples of Networks with Negative Coupling}

Now we consider examples in which there are negative connections,
i.e., some of the entries of the adjacency matrix are negative,
$A_{nm} < 0$. In our next example, we construct first an undirected
network with $N = 1500$ nodes and $400$ connections per node. We
then set $A_{nm} = 0$ if $n$ and $m$ are not connected, and if they
are we set $A_{nm}$ to $1$ with probability $q$ and to $-1$ with
probability $1-q$.

First we consider the case $q = 2/3$, so that one third of the
connections are negative [example (iii)]. In Fig.~\ref{inhi} we
compare the numerical solution of Eq.~(\ref{eq:coupled}) with our
theoretical approximations in Eqs.~(\ref{eq:betaprime2}) and
(\ref{eq:fdainhi}) for ten realizations of the network and
frequencies. We show the average of the order parameter $r^2$ over
$10$ realizations of the network (triangles with thin error bars),
the average of the TAT [Eq.~(\ref{eq:betaprime2}), solid line with
oval error bars], and the average of the FDA
[Eq.~(\ref{eq:fdainhi}), dashed line]. The error bar widths
represent one standard deviation of the sample of $10$ realizations.
As in the previous examples, the FDA did not show noticeable
variations for different realizations of the network. We observe
that the order parameter computed from our theory yields a slightly
larger value than that obtained from the numerical solution of
Eq.~(\ref{eq:coupled}), but in general both the transition point and
the behavior of the order parameter are described satisfactorily by
the theory.

In this case, the phases $\psi_n$ obtained from numerical solution
of Eq.~(\ref{eq:betaprime2}) do not depend on $n$, i.e., $\psi_n =
\psi_m$ for all $n$, $m$. This can be understood on the basis that
there are not enough negative coupling terms to make the right hand
side of Eq.~(\ref{eq:betass}) negative, so that a solution exists in
which all the phases $\psi_n$ are equal. As mentioned in
Sec.~\ref{inhibitory}, the difference in the phases in
Eq.~(\ref{eq:betaprime2}) prevents the right hand side of
Eq.~(\ref{eq:betass}) from becoming negative in the presence of
negative connections. As a confirmation of this we note that as
$k\to \infty$ the order parameter $r$ appears to  approach $1/3$
(the dotted-dashed horizontal line in Fig.~\ref{inhi}), which
corresponds to $(\psi_n - \psi_m)\to 0$ in Eq.~(\ref{eq:cosfi2}) for
$q = 2/3$. The fact that both the phases $\psi_n$ and $\theta_n$ do
not depend on $n$ as $k \to \infty$ is consistent with
Eq.~(\ref{eq:locked}).

\begin{figure}[h]
\begin{center}
\epsfig{file = 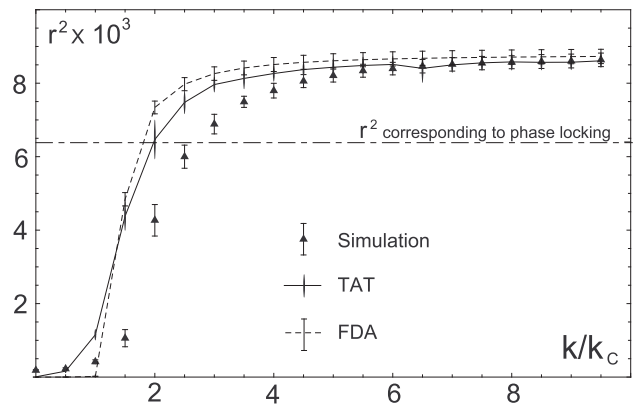, clip =  ,width=1.0\linewidth}
\caption{
Average of the order parameter $r^2$ obtained from numerical solution of Eq.~(\ref{eq:coupled})
over $10$ realizations of the network with $q = 0.54$ and frequencies (triangles)
and average of the TAT (solid line) as a function of $k/k_c$.
Note the different scale in the horizontal axis as compared with the previous figures.
The horizontal dotted-dashed line represents the value of the order parameter if the oscillators
were phase locked ($\theta_n = \theta_m$ for all $m$ and $n$).}
\label{inhi54}
\end{center}
\end{figure}

In order to consider a case in which the effect of the negative
connections is more extreme, we consider a network constructed as
described above with with $q = 0.54$  [example (iv)]. In
Fig.~\ref{inhi54} we compare the numerical solution of
Eq.~(\ref{eq:coupled}) with our theoretical approximations in
Eqs.~(\ref{eq:betaprime2}) and (\ref{eq:fdainhi}) for ten
realizations of the network and frequencies. We show the average of
the order parameter $r^2$ over $10$ realizations of the network
(triangles with thin error bars), the average of the FDA [Eq.~(\ref{eq:fdainhi}),
dashed line with thin error bars] and the average of the TAT
[Eq.~(\ref{eq:betaprime2}), solid line with oval error bars]. When
numerically solving Eq.~(\ref{eq:betaprime2}) by iteration of
Eq.~(\ref{eq:betaprime3}), on some occasions a period two orbit was
found instead of the desired fixed point. If we denote the left hand
side of Eq.~(\ref{eq:betaprime3}) by $z^{j+1}_n$ and the right hand
side by $f(z^{j}_n)$, we found that convergence to a fixed point was
facilitated by replacing the right hand side by $[z^{j}_n +
f(z^{j}_n)]/2$ and finding the fixed points of this modified system.

In this example, at low coupling strengths [roughly $k/k_c \lesssim
4$, where $k_c$ is computed from Eq.~(\ref{kcinhi})] the order
parameter computed from numerical solution of Eq.~(\ref{eq:coupled})
is smaller than that obtained from the TAT and FDA. As $k$ increases,
however, the TAT and FDA theories captures the asymptotic value of the order
parameter $r$. We note that in this case the asymptotic value is
larger than that corresponding to phase locking [i.e., the one
obtained by setting $\psi_n = 0$ in Eq.~(\ref{eq:cosfi2}), $r
\approx 0.54 -0.46 = 0.08$], which we indicate by a horizontal
dotted-dashed line in Fig~\ref{inhi54}, and much smaller than $r =
1$, the value corresponding to no frustration [i.e., $\psi_n -
\psi_m = 0$ for $A_{nm} > 0$ and $\pi$ for $A_{nm} < 0$ in
Eq.~(\ref{eq:cosfi2})]. The small scale of the horizontal axis is
due to the fact that we are plotting $r^2$, and to our definition of
the order parameter which assigns a value of $1$ to a non frustrated
configuration. The small value of the order parameter indicates a
strong frustration.

\begin{figure}[h]
\begin{center}
\epsfig{file = 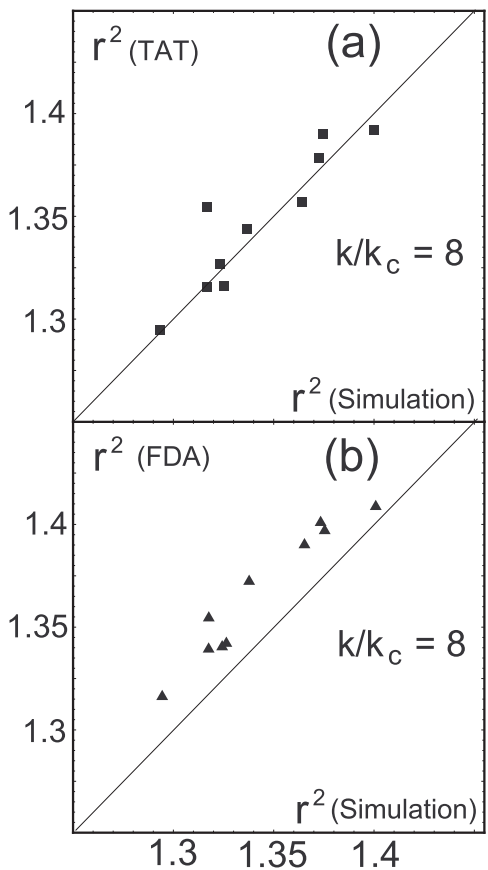, clip =  ,width=0.8\linewidth}
\caption{
Order parameter $r^2$ obtained from (a) the TAT (triangles) and from (b) the FDA
(boxes) versus the value obtained from numerical solution of Eq.~(\ref{eq:coupled})
for $k/k_c = 8$. The solid line is the identity.
Besides a small positive bias in the FDA, the theories track the spread in the results of
the numerical solution for different realizations.
}
\label{figcorre}
\end{center}
\end{figure}

We note that in this example, in contrast with the examples discussed so far,
there is variation in the values of the order parameter predicted by the FDA
for different realizations of the network. This indicates that,
as the expected value of the coupling strengths $A_{nm}$ becomes small
(i.e., $|q - 1/2|$ small), fluctuations
due to the realization of the network become noticeable. Although
the values predicted by the FDA and TAT depend on the realization of the
network and frequencies, we note for $k/k_c \gtrsim 6$ that these values track
the values observed for the numerical simulations
of the corresponding realization. As an illustration of this, we plot in Fig.~\ref{figcorre}(a)
the values of $r^2$ obtained from the TAT (triangles) and in Fig.~\ref{figcorre}(b)
the values of $r^2$ obtained from the FDA (boxes) versus the value obtained
from numerical solution of Eq.~(\ref{eq:coupled})
for $k/k_c = 8$. Besides a small positive bias in the FDA, the theories track the spread
in the results of the numerical solution for different realizations. Some bias in the FDA is not
surprising, because we averaged the right hand side of the nonlinear equation (\ref{eq:betass})
for the TAT in order to get Eq.~(\ref{eq:betaint}) for the FDA. Nonetheless, the bias is
extremely small in most of our examples.

The behavior observed in Fig.~\ref{inhi54} at $k/k_c \lesssim 4$ can be
interpreted as a shift in the transition point to a larger value of
the coupling strength, and is reminiscent of what occurs when the
time fluctuations  [$k h_n(t)$ in Eq.~(\ref{eq:coupled0})] neglected
in Eq.~(\ref{eq:coupled2}) have an appreciable effect \cite{onset}.
We believe that the time fluctuations have a more pronounced effect
as the number of negative connections becomes comparable to the
number of positive connections (i.e., as $|q-1/2|$ becomes small)
because the critical coupling strength $k_c$ becomes large
(roughly $k_c \sim |q -1/2|^{-1}$).
In particular, with positive connections, the condition for
neglecting $k h_n(t)$ was that the number of connections to each node
was large. In contrast, for the present case, the analogous
statement would be that $|q-1/2|$ times the number of connections is
large, which is much less well satisfied, $|q - 1/2|400 =
0.04\times400 = 16$. The extreme case of zero mean coupling has
already been studied numerically by Daido \cite{daido2}, who found
that in this case the oscillators lock in the sense that their
average frequency is the same, but their phases diffuse. As argued
in Ref.~\onlinecite{onset}, such fluctuations have the effect of shifting the
transition to larger values of the coupling strength. It would be
interesting to carry on simulations in networks with a much larger
number of connections per node, as the effect of fluctuations would
likely be reduced.

We also considered a case in which the adjacency matrix is
asymmetric and has mixed positive/negative connections. For $N =
1500$ nodes, we constructed an adjacency matrix by setting its
nondiagonal entries to $1$, $-1$, and $0$ with probability $8/45$,
$4,45$, and $11/15$, respectively. The latter probability yields an
expected number of connections of $400$. Our theories work
satisfactorily in this case, and, since the results are similar to
those in Fig.~\ref{inhi}, we do not show them. In this case there is
no guarantee that there is a real eigenvalue [as needed for
estimating the critical coupling strength in Eq.~(\ref{eq:kc})], or
that the largest real eigenvalue (if there is one) has the largest
real part. Numerically, we find that for matrices constructed as in
this example there is a real positive eigenvalue and that,
furthermore, it is well separated from the largest real part of the
remaining eigenvalues (see Fig.~\ref{eithe}). We also find this for
other values of $q$ provided $|q - 1/2|$ is not too small. We
provide a discussion of this issue and show the spectrum of the
adjacency matrix in Appendix \ref{appendixa}.

\section{Discussion}\label{discussion}

In this paper, we have considered interacting phase oscillators
[Eq.~(\ref{eq:coupled})] connected by directed networks and networks
with mixed positive/negative connections.

The previous theory of Ref.~\onlinecite{onset} given by
Eq.~(\ref{eq:betass}) (the time averaged theory, TAT) can still be
applied for asymmetric networks with purely positive coupling and
was found to give good predictions, applicable to
{\it individual} asymmetric random realizations [Figs.
\ref{fig:figpeasy}(b), \ref{fig:figpe110}(b)]. The previous theory
given by Eq.~(\ref{eq:betaint}) (the frequency distribution
approximation, FDA) can also still be applied for asymmetric
networks with purely positive coupling and was found to give good
predictions applicable to the {\it ensemble average behavior}
of asymmetric network realizations [Figs. \ref{fig:figpeasy}(a),
\ref{fig:figpe110}(a)]. The perturbative theory for the FDA was
generalized to account for directed networks
[Eqs.~(\ref{eq:secondpt}) and (\ref{eq:eta1dire})], as was the
previous undirected network mean field theory, MFT (generalized from
Eqs.~(\ref{eq:betasumed})-(\ref{eq:eta2}) to
Eqs.~(\ref{mftdire})-(\ref{eq:eta2dire})]. In our example (ii),
which had a very strong asymmetry, we found that our directed FDA
perturbation theory [Eqs.~(\ref{eq:secondpt}) and
(\ref{eq:eta1dire})] gave a good description of synchronization, but
that the directed mean field approximation gave a transition to
synchronization at a coupling substantially below that observed. In
contrast, for example (i), in which the coupling matrices were
individually asymmetric but their ensemble average was symmetric,
the mean field theory (and all the other theories in
Sec.~\ref{dire}) gave good results.

For the case of mixed positive/negative couplings we presented a
generalization of the TAT and FDA,
Eqs.~(\ref{eq:betaprime2})-(\ref{kcinhi}). We tested these results
on two examples, example (iii) in which a fraction $1-q = 1/3$ of
the couplings were negative, and example (iv) in which a fraction
$1-q = 0.46$ of the couplings were negative. For example (iii) we
found that iteration of Eq.~(\ref{eq:betaprime3}) converges to a
fixed point with $\psi_n -\psi_m =0$, and thus the result is similar
to the case where all connections are positive. In example (iv), the
result of iteration of Eq.~(\ref{eq:betaprime3}) yields nontrivial
values for the phases $\psi_n$. In this case we found good agreement
between the solutions of (\ref{eq:coupled}) and the theory for the
order parameter for $k/k_c$ large ($k/k_c \gtrsim 4 $), but that for
smaller $k/k_c$ ($k/k_c \lesssim 4 $), although yielding
qualitatively similar behavior to that observed (Fig.~\ref{inhi54}),
the theory overestimates the order parameter. Analogous to similar
observations for symmetric networks with only positive coupling
\cite{onset}, we speculate (Sec.~\ref{examples}) that this is a
finite size effect associated with the fact that the effective
number of connections given in this example by $|q-1/2|400 = 16$ is
not sufficiently large to justify neglect of $k h_n(t)$ in
Eq.~(\ref{eq:coupled0})

In order to isolate the effect of the asymmetry and the negative
connections, we considered networks in which the degree distribution
is very narrow. The combined effect of these factors with different
heterogeneous degree distributions (e.g., scale free networks
\cite{barabasi2}) and with correlations in the network (in
particular, degree-degree correlations) is still open to
investigation.

In practice, one could be interested in networks in which the
asymmetry in the connections is strongly correlated with the sign of
the coupling (in analogy to some models in neuroscience
\cite{wang}). Although we did not study such a case here, we believe
our theory provides a good starting point to study the emergence of
synchronization in these kind of structured complex networks.

\appendix

\section{}\label{appendixa}

In this Appendix we discuss the characteristics of the spectrum of
the adjacency matrices considered in our examples. Although we will
focus here on asymmetric matrices, a similar argument works for
symmetric matrices. The matrices we consider are relatively sparse,
with the position of the nonzero entries being chosen randomly
(e.g., in the symmetric case, the position of the nonzero entries is
chosen when constructing the network using the configuration model),
and their values being also determined randomly
from a given probability distribution (e.g., $1$ with probability
$q$ and $-1$ with probability $1-q$). Our interest is focused on the
gap between the largest real eigenvalue (if there is one) and the
largest real part of the other eigenvalues. In Ref.~\onlinecite{timme} the
spectrum of certain large sparse matrices with average eigenvalue
$0$ and row sum $\sum_{m=1} A_{nm}= 1$ was described and a heuristic
analytical approach was proposed. Using results for matrices with
zero mean Gaussian random entries \cite{sommers}, Ref. \onlinecite{timme}
predicts that the spectrum of the non-Gaussian random matrices they
consider consists of a trivial eigenvalue $\lambda = 1$ with the
remaining eigenvalues distributed uniformly in a circle centered at
the origin of the complex plane with radius
\begin{equation}\label{a1}
\rho = \sqrt{N} \sigma,
\end{equation}
where $\sigma^2$ is the variance of the entries of the matrix.
\begin{figure}[h]
\begin{center}
\epsfig{file = 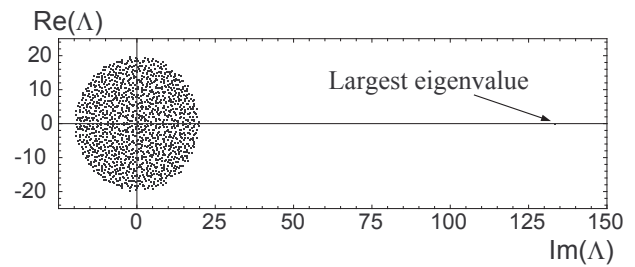, clip =  ,width=1.0\linewidth}
\caption{
Complex eigenvalues $\Lambda$ (dots) of a $1500\times 1500$ random matrix whose off diagonal entries
are $1$, $-1$ or $0$ with probabilities $8/45$, $4/45$, and $11/15$, respectively. One eigenvalue is located
at $\lambda = 136.2$, while the other $1499$ eigenvalues uniformly fill a circle of radius $\rho$ centered
at the origin of the complex $\Lambda$ plane. Note that $\rho \approx 19.8$ is substantially less than
$136.2$. Comparing with the theory in the Appendix, Eq.~(\ref{a2}) yields a prediction of $133.3$ for the maximum
real eigenvalue while Eq.~(\ref{a1}) predicts $19.7$ for $\rho$.
These are in excellent agreement with our numerically determined values.
}
\label{eithe}
\end{center}
\end{figure}
We find that this approach also succeeds in describing the spectrum
of the matrices in our examples. In our case, the diagonal entries
are $0$, so that the average eigenvalue is also $0$ as in Ref.~\onlinecite{timme}.
We find that there is always a largest real eigenvalue
approximately given by the mean field value
\begin{equation}\label{a2}
\lambda = \langle {\tilde{d}}^2\rangle/\langle \tilde{d}\rangle
\end{equation}
(see Refs.~\onlinecite{onset, chung}), where $\tilde{d}_n = \sum_{m=1}^N A_{nm}$
and $\langle \tilde{d}^2 \rangle = \sum_{n=1}^N \tilde{d}_n^2 $,
which in the case considered in Ref.~\onlinecite{timme} reduces to $\lambda =
1$. We also numerically confirm that the remaining eigenvalues are
uniformly distributed in a circle of radius $\rho$ as described in Ref.~\onlinecite{timme}.
This is illustrated in Fig.~\ref{eithe}.

Thus for $N \gg 1$ if $\lambda > \rho$ there is a gap of size
$\lambda - \rho$ between the largest real eigenvalue and real part
of the rest of the eigenvalue spectrum. Using Eqs.~(\ref{a1}) and
(\ref{a2}) it can be shown that, for networks with large enough
number of connections per node or with enough positive (or negative)
bias in the coupling strength, there is a wide separation between
the largest eigenvalue and the largest real part of the remaining
eigenvectors. For symmetric matrices, similar results apply (i.e,
the bulk of the spectrum of the matrix $A$ can be approximately
obtained as described above using Wigner's semicircle law).


\begin{thebibliography}{99}





\bibitem{pikovsky} A. Pikovsky, M.G. Rosenblum, and J. Kurths,
{\it Synchronization: A universal concept in nonlinear sciences},
(Cambridge University Press, Cambridge, 2001).

\bibitem{mosekilde} E. Mosekilde, Y. Maistrenko, and D. Postnov,
{\it Chaotic Synchronization: Applications to Living Systems} (World Scientific,
Singapore, 2002).

\bibitem{pecora2} L.M. Pecora and T.L. Carroll, Phys. Rev. Lett. {\bf 80}, 2109 (1998).

\bibitem{pecora3} M. Barahona and L.M. Pecora, Phys. Rev. Lett. {\bf 89}, 054101 (2002).

\bibitem{takashi} T. Nishikawa, A. E. Motter, Y.-C. Lai, and F. C. Hoppensteadt,
Phys. Rev. Lett. {\bf 91}, 014101 (2003).

\bibitem{jadbabaie}
A. Jadbabaie, N. Motee, and M. Barahona. Proceedings of the American Control Conference (ACC 2004).

\bibitem{bahiana}
J.~L. Rogers and L. T. Wille, Phys. Rev. E {\bf 54}, R2193 (1996);
M. S. O. Massunaga and M. Bahiana, Physica D {\bf 168-169}, 136, (2002);
M. Mar{\'o}di, F. d'Ovidio, and T. Vicsek, Phys. Rev E {\bf 66}, 011109 (2002).

\bibitem{moreno} Y. Moreno and A. E. Pacheco, Europhys. Lett. {\bf 68}, 603
(2004).

\bibitem{ichinomiya} T. Ichinomiya, Phys. Rev. E {\bf 70}, 026116 (2004).

\bibitem{lee} D.-S. Lee, eprint cond-mat/0410635.

\bibitem{ichinomiya2} T. Ichinomiya, eprint cond-mat/0507285.

\bibitem{onset} Juan G. Restrepo, Brian R. Hunt, and Edward Ott,
Phys. Rev. E {\bf 71}, 036151 (2005).

\bibitem{kuramoto} Y. Kuramoto, {\it Chemical Oscillations, Waves, and Turbulence},
(Springer-Verlag, Berlin, 1984).

\bibitem{strogatz} S. H. Strogatz, Physica D {\bf 143}, 1 (2000).

\bibitem{newman1} M.E.J. Newman, SIAM Review {\bf 45}, 167 (2003).

\bibitem{barabasi1}
A.-L. Barab\'{a}si, and R. Albert, Rev. Mod. Phys. {\bf 74}, 47 (2002).

\bibitem{footnote} In Ref.~\onlinecite{onset} we argue, based on
Ref.~\onlinecite{lee}, that Eqs.~(\ref{perturba}) and (\ref{eq:eta1}) are valid
in the limit $N\to \infty$ for degree distributions $p(\tilde{d}) $
such that $\int_1^\infty p(\tilde{d})\tilde{d}^4d\tilde{d}$ is
finite.

\bibitem{bapat} R. B. Bapat and T. E. S. Raghavan,
{\it Nonnegative Matrices and Applications} (Cambridge University
Press, Camdridge, 1997).

\bibitem{daido2} H. Daido, Phys. Rev. Lett. {\bf 68}, 1073 (1992).

\bibitem{takashi2} T. Nishikawa, Y.-C. Lai, and F. C. Hoppensteadt,
Phys. Rev. Lett. {\bf 92}, 108101 (2004).



\bibitem{barabasi2} A.-L. Barab\'{a}si, and R. Albert, Science {\bf 286}, 509 (1999).

\bibitem{wang} X.-J. Wang, Neuron {\bf 36}, 955 (2002).

\bibitem{timme} M. Timme, F. Wolf, and T. Gheisel, Phys. Rev, Lett. {\bf 92}, 074101 (2004).

\bibitem{sommers} H. J. Sommers, A. Crisanti, H. Sompolinsky, and Y. Stein,
Phys. Rev. Lett. {\bf 60}, 1895 (1988).

\bibitem{chung} F. Chung, L. Lu and V. Vu, Proc. Natl. Acad. Sci., {\bf 100}, 6313  (2003).

\end{thebibliography}
\end{document}